# Spin and charge density waves in the quasi-one-dimensional $KMn_6Bi_5$

Jin-Ke Bao[1,2,3], Huibo Cao[4], Matthew J. Krogstad,[1,5] Keith M. Taddei[4], Chenfei Shi[2], Shixun Cao[2,3], Saul H. Lapidus[5], Sander van Smaalen[6], Duck Young Chung[1], Mercouri G. Kanatzidis[1,7], Stephan Rosenkranz[1], Omar Chmaissem[1,8,*]

[1]*Materials Science Division, Argonne National Laboratory, Lemont, Illinois 60439, USA*
[2]*Department of Physics, Materials Genome Institute and International Center for Quantum and Molecular Structures, Shanghai University, Shanghai 200444, China*
[3]*Shanghai Key Laboratory of High Temperature Superconductors, Shanghai University, Shanghai 200444, China*
[4]*Neutron Scattering Division, Oak Ridge National Laboratory, Oak Ridge, Tennessee 37831, USA*
[5]*Advanced Photon Source, Argonne National Laboratory, Lemont, Illinois 60439, USA*
[6]*Laboratory of Crystallography, University of Bayreuth, Bayreuth 95447, Germany*
[7]*Department of Chemistry, Northwestern University, Evanston, Illinois 60208, USA*
[8]*Department of Physics, Northern Illinois University, DeKalb, Illinois 60115, USA*

*Corresponding author: chmaissem@anl.gov

**Abstract**

$A$Mn$_6$Bi$_5$ materials ($A$ = Na, K, Rb and Cs) consisting of unique Mn-cluster chains emerge as a new family of superconductors with the suppression of their antiferromagnetic (AFM) order under high pressures. Here, we report transverse incommensurate spin density waves (SDWs) for the Mn atoms with a propagating direction along the chain axes as a ground state for KMn$_6$Bi$_5$ by single crystal neutron diffraction. The SDWs have a refined amplitude of ~2.46 $\mu_B$ for the Mn atoms in the pentagons and ~0.29 $\mu_B$ with a large standard deviation for Mn atoms in the center between the pentagons. AFM dominate both the nearest-neighbor Mn-Mn interactions within the pentagon and next-nearest-neighbor Mn-Mn interactions out of the pentagon (along the propagating wave). The SDWs exhibit both local and itinerant characteristics probably formed by a cooperative interaction between local magnetic exchange and conduction electrons. A significant magnetoelastic effect during the AFM transition, especially along the chain direction, has been demonstrated by temperature-dependent x-ray powder diffraction. Single crystal x-ray diffraction below the AFM transition revealed satellite peaks originating from charge density waves along the chain direction with a **q**-vector twice as large as the SDW one, pointing to a strong real space coupling between them. Our work not only manifests a fascinating interplay among spin, charge, lattice and one dimensionality to trigger intertwined orders in $KMn_6Bi_5$ but also provides important piece of information for the magnetic structure of the parent compound to understand the mechanism of superconductivity in this new family.

**Introduction**

In contrast to conventional superconductivity with s-wave electron Cooper pairs predicted by Bardeen-Cooper-Schrieffer theory [1], unconventional superconductivity has repeatedly manifested itself by electronic pairing mediated by suppression of the magnetic order of a candidate magnetic material via chemical doping or by the application of external physical pressure [2]. The underlying mechanisms responsible for the induction of unconventional superconductivity remain mysterious despite groundbreaking discoveries of numerous superconducting families including the famous cuprates [3], iron pnictides [4], heavy fermion systems [5] and organic Bechgaard salts [6].

Superconductivity was recently reported in a series of Mn-based quasi-one-dimensional (Q1D) $AMn_6Bi_5$ ($A$ = K, Rb and Cs) materials [7–9] via the suppression of their antiferromagnetic (AFM) orders under 9-15 GPa external pressures [9–11]. Considering the similar phase diagrams of unconventional superconductivity tuned by high pressure [2,12–14], it is reasonable to classify $AMn_6Bi_5$ as another class of unconventional superconductors whose underlying physics needs to be uncovered. Moreover, quasi-one-dimensionality, weakly coupled one-dimensional unit where electron-electron correlations and quantum fluctuations become enhanced due to the quantum confinement of electrons [15], enriches the physics in this family. Intriguing collective electronic instabilities or behaviors, for example, Tomonaga-Luttinger liquid with spin-charge separation [16], charge density wave (CDW) [17,18], spin density wave (SDW) [19,20] and unconventional superconductivity with spin-triplet pairing [21–26], have been either predicted or observed in Q1D systems.

$A$Mn$_6$Bi$_5$ ($A$ = Na, K, Rb and Cs) is a model system that offers a rare opportunity for investigations of electronic instabilities in a nuclear structure which consists of unique [Mn$_6$Bi$_5$]$^-$ nanowires with counterions $A$ intercalated in between to form a Q1D crystal structure [7,8,27], see Figure 1(a,b). One-dimensional intra and inter-nanowire motific defects originating from the Q1D bonding characteristics has also been visualized by state-of-the-art scanning transmission electron microscope in KMn$_6$Bi$_5$ [28]. $A$Mn$_6$Bi$_5$ compounds exhibit AFM properties with transition temperatures ranging between 47 and 82 K depending on the size of the $A$-site alkaline ions which tunes the distance between the nanowires and, thus, the interchain magnetic coupling as well as the local geometry and, thus, the intrachain magnetic exchange coupling.

Clearly, understanding the physics of unconventional superconductors requires an intimate knowledge of the exact nature and tuning properties of their magnetic ground states as they compete with induced superconductivity. The magnetic ground state of this emerging family of $A$Mn$_6$Bi$_5$ superconductors is yet to be determined experimentally. For RbMn$_6$Bi$_5$ [8], recent first-principles calculations suggested that the energetically favorable AFM ground state is possibly helical with a reduced Mn magnetic moment. In this work, we use neutron single crystal diffraction to demonstrate the formation of a complex incommensurate magnetic structure for KMn$_6$Bi$_5$ consisting of multiple AFM spin density waves with a period of about twelve unit-cells (~55.4 Å) propagating along the short $b$-axis. Furthermore, using synchrotron x-ray single crystal diffraction, we observed a CDW with a propagation vector twice as large as that of the SDW.

**Experimental**

The preparation methods of samples for neutron and x-ray diffraction are described in Supplemental Material (SM) [ [29]]. Synchrotron powder and single crystal x-ray diffraction were performed at beamline 11-BM-B and 6-ID-D of the Advanced Photon Source at Argonne National Laboratory (ANL), respectively. Magnetic structure determination and order parameter scans were performed using neutron powder diffraction at beamline HB-2A and neutron single crystal diffraction at beamline HB-3A [30] at the high flux isotope reactor (HFIR) at Oak Ridge National Laboratory (ORNL). Data collection details, processing and analysis are provided in the SM [29]. Crystal and magnetic structures were plotted using the software VESTA [31].

**Results and discussion**

*Structural Properties*

$KMn_6Bi_5$ crystallizes in a monoclinic crystal structure (No. 12, *C*2/*m*) consisting of two $[Mn_6Bi_5]^-$ in one unit cell [7], see Figure 1(a). Each $[Mn_6Bi_5]^-$ nanowire can be decomposed into a chain of corner-shared Mn-centered icosahedral clusters (13 Mn atoms) encapsulated inside a Bi nanotube, see Figure 1(b). The Mn clusters consist of five independent Mn sites (Mn1 – Mn5) that form two flat pentagons lying within the *ac* planes, at *y*-coordinates of 0 and 0.5, with an additional independent Mn6 site (referred to as central Mn) occupying central positions, at *y*-coordinates of 0.25 and 0.75, between any two adjacent pentagons stacked along the *b*-axis. The Mn pentagons are structurally related via the symmetry operators of the monoclinic *C*2/*m* space group. No sign of any structural symmetry transformation was detected across its AFM transition from temperature-dependent x-ray powder diffraction measurements. The temperature-dependent behavior of the diffraction data agrees with an overall positive thermal expansion of the material and the anomalies at ~75 K (the opposite responses of the peaks (10 0 –3) and (0 2 0) during the transition) imply significant anisotropic magnetoelastic coupling between its crystal lattice and magnetic order [32,33], Figure 1(c). Additional peaks are tracked as shown in Figure S1 [29].

Quantitative temperature-dependent lattice parameters of $KMn_6Bi_5$ determined from Rietveld refinements are shown in Figure 1(d). With the exception of a subtle slope change of the unit cell volume, all the lattice parameters display a clear anomaly at ~75 K. It is worth emphasizing the rapidly decreasing lattice parameter *b* after the system enters the AFM state, which is ascribed to strong AFM exchange magnetostriction [34] within the Mn cluster chains as described below. In contrast, the lattice parameters *a* and *c* both exhibit a smaller increase just below the paramagnetic to antiferromagnetic phase transition.

Given that the interchain magnetic interactions are much weaker than their intrachain counterparts, one can argue that the negative thermal expansion displayed by *a* and *c* results from elastic strains produced by compression of the lattice along the chain direction (*i.e.*, the *b*-axis). On the other hand, the relatively larger positive thermal expansion along the *b*-axis, Figure 1(d), suggests that one-dimensional magnetic interactions along the chain direction play a dominant role in determining the overall magnetoelastic properties of the material below 99 K.

*Incommensurate Spin Density Waves (ISDW)*

Neutron diffraction from a $KMn_6Bi_5$ single crystal was utilized to determine the magnetic structure. The high quality of our single crystal is demonstrated by the excellent goodness-of-fit and agreement factors obtained at 4 K for the refined nuclear structure as shown in Figure S2(a) and Table S1 in the SM [29]. Reciprocal space scans revealed the existence of a new set of reflections of magnetic origin at temperatures below 75 K, which we successfully indexed using an incommensurate propagation vector $\mathbf{q_s}$ = (1, 0.418, 0). This result is markedly different from any of the theoretical commensurate magnetic models proposed by Chen *et al.* [8]. The order parameter scan using the magnetic reflection (1 0.418 1) gave a fitted critical component $\beta$ of 0.15(3) for $KMn_6Bi_5$ (Figure S2(c) in SM [ [29]), typical of systems with reduced dimensionality and much smaller than the theoretical value of 0.37 for three-dimensional Heisenberg systems [35], which agrees well with the proposed Q1D magnetic interactions [36,37].

With a total of 6 independent Mn sites producing a magnetic sublattice with 24 Mn ions per unit cell, symmetry-breaking magnetic interactions are expected within and between the Mn clusters. A broken *C*-centering symmetry results in twelve independent Mn sites as shown in Figures 2 and 3 and described in the magnetic modeling in SM [29]. In our final model with a goodness-of-fit agreement factor of about 8% (Figure S2(b)), amplitude modulated SDWs with a period of about twelve-unit cell long (~55.4 Å) along *b* axis are observed through all the Mn atoms, see Figure 2. Additional 2-fold rotation symmetry along the (0.5, *y*, 0.5) axis transforms the magnetic moments (*u*, 0, *w*) of Mn atoms in one Mn-cluster chain to (-*u*, 0, -*w*) of Mn atoms in another Mn-cluster chain as shown in Figure 3. Each single wave contains one Mn-atom chain with a Mn-Mn distance equal to that of the lattice parameter of *b*. Two separate waves are mapped into one single Mn-atom chain for Mn6 and Mn12 atoms with the Mn-Mn distance decreasing by half. We note that the waves passing through the basal pentagons (*i.e.,* at *y* = 0) are ahead of the waves that propagate through the Mn pentagons at *y* = 0.5 by about 284°. This phase shift determined by the strong coupling of SDWs between two neighboring Mn pentagons within the same Mn-cluster chain is the major driving force leading to breaking the *C*-centering of the parent space group. Other models such as spiral helical magnetic ones were tried but turned out to be unsuccessful or converged to the current model, see refinement details in SM [29]. Our model produces an excellent fit to the magnetic intensities observed by neutron powder diffraction data collected at 3 K (Figure S3).

There are 12 independent SDWs of Mn atoms coupled and propagated together along one single Mn-cluster chain in $KMn_6Bi_5$. The magnetic configuration can be described by the general formula [38]:

$$\mathbf{m_{ji}} = \mathbf{m_i} \cos(-2\pi \mathbf{q_s} \cdot \mathbf{R_j} + \varphi_i) \ (i = 1\text{-}24)$$

Where $\mathbf{m_{ji}}$ and $\mathbf{m_i}$ with a vector form of (*u*, 0, *w*) are the magnetic moments in the *j*th unit cell and wave amplitudes for the Mn*i* atom, respectively, $\mathbf{R_j}$ is the translation vector, $\mathbf{q_s}$ is the propagation vector and $\varphi_i$ is the phase for one certain Mn atom. The magnetic moments and directions of Mn6 and Mn12 cannot be determined accurately from our current dataset due to the major magnetic contribution from the Mn pentagons with significantly larger magnetic moments. For clarity, the waves from different Mn pentagons are separated as shown in Figure 3(a,b). Refined magnetic structure parameters are presented in Table S2 in SM [29].

All the SDWs along the chain are transverse in $KMn_6Bi_5$ since the propagating direction is perpendicular to the spin direction. In a conventional transverse SDW, the modulated magnetic moments point in one direction for half of the wave then reverse direction in the other half as in Cr metal [39], for example. In this work, the component of the vector $\mathbf{q_s}$ along the chain direction is 0.418 which corresponds to a phase shift of 150° when moving one unit cell along the chain direction. As a result, the magnetic moments are not only modulated as expected for a transverse SDW but also coupled antiferromagnetically along the wave direction.

SDW instability can arise from a low-dimensional metallic system due to electron-electron interactions [40], which fully matches the scenario in metallic Q1D $KMn_6Bi_5$ with a Mn-3d system. Indeed, carrier concentration decreases during the AFM transition probably ascribed to the partial gapping of the Fermi surface [7]. A simple calculation of the oxidation state reveals that a total of +14 charges must be accounted for by the six Mn ions assuming $K^+$ and $Bi^{3-}$ [41]. Our refinements of ~2.46 $\mu_B$ and 0.29 $\mu_B$ as the waves magnitudes and their corresponding root mean square values of 1.73 and 0.21 $\mu_B$ for the pentagon and central Mn atoms, respectively, suggest possible charge disproportionation and ordering of the different magnetic moments with five $Mn^{2+}$ (5 $\mu_B$ for high spin) making the pentagons and one $Mn^{4+}$ (1 $\mu_B$ for low spin) along the pentagon axes. The reduced magnetic moment of Mn in $KMn_6Bi_5$ can be realized by delocalization of the 3d electrons from Mn-Mn metallic bonding within the chain [7]. The significantly small magnetic moment in the central Mn ions is supported by the very short Mn-Mn distance (~ 2.28 Å at 4 K) where strong metallic bonding between Mn atoms exists [42]. First-principles calculations on $AMn_6Bi_5$ do reveal that Mn-3d electrons dominate the density of states around the Fermi level, proving the itinercy of the 3d electrons [8,27]. The small magnetic moments obtained in the model are also basically consistent with a small effective magnetic moment per Mn atom obtained from the Curie-Weiss fitting in the experiment [7,10]. The disproportionation of magnetic moments of Mn atoms in $KMn_6Bi_5$ is reminiscent of the coexistence of magnetic (2-3 $\mu_B$) and almost nonmagnetic (0.2-0.6 $\mu_B$) Mn atoms in the allotrope $\alpha$-Mn [43]. Our results agree well with DFT calculations performed by Chen *et al.* [8] from which the central Mn ions are found to have a much smaller magnetic moment than the pentagon Mn. The small magnetic moments in $KMn_6Bi_5$ may be key to achieving superconductivity by high pressures like CrAs [12] and MnP [13] where fragile AFM can be easily quenched.

While magnetic moments from central Mn atoms (0.21 $\mu_B$) are similar to that of a typical SDW system [3,6,44], the magnetic moments (1.73 $\mu_B$) of Mn in the pentagon in $KMn_6Bi_5$ are larger, indicating that magnetism comes from both localized and itinerant electrons as in GdSi, for example, where local magnetic moments and itinerant electrons form a cooperative magnetic configuration [45]. In terms of a local magnetic picture, the Mn-Bi-Mn magnetic exchange geometry in $KMn_6Bi_5$ is similar to the ones in a prototypical insulator $BaMn_2Bi_2$ where AFM coupling dominates for both nearest and next-nearest neighboring Mn-Mn interactions [46]. Canted AFM configuration for two nearest-neighbor Mn atoms in the pentagon in $KMn_6Bi_5$ is a compromise arising from geometric frustration and AFM exchange coupling via Bi atoms in the same plane, see Figure 2. AFM coupling in the waves for Mn pentagons can be understood by next-nearest-neighbor Mn-Mn magnetic exchange using Bi atoms sandwiched between them as a bridge [8]. In contrast to other pressure-induced superconductors containing 3d electrons such as CrAs [12] and MnP [13] with some degree of ferromagnetic interactions, AFM dominates in $KMn_6Bi_5$ with a one-dimensional and strongly coupled magnetic sublattice, which poses a constraint on creating a theoretical model to understand its superconductivity under high pressure.

*Incommensurate Charge Density Waves (ICDW)*

Single crystal synchrotron x-ray diffraction revealed the existence of satellite reflections, in addition to the main Bragg peaks, at 30 K, well below the magnetic phase transition, as shown in Figure 4. These extra peaks are successfully indexed using a propagation vector $q_c \approx (0, 0.83, 0)$, proving the existence of CDW along the chain direction. The charge propagation vector $q_c$ is almost double that of the incommensurate modulated component of the SDW vector $q_s$ indicating their strongly coupling in real space as also observed for example in Cr [39,47], underdoped cuprates [48] and nickelates [49]. *C*-centering symmetry in the crystal structure is still preserved at 30 K based on the rule of the extinction condition for the main reflections. Integrated synchrotron intensities at 30 and 297 K are shown in Figure S4 and S5 in SM [29].

Compared with SDW, CDW is more subtle in $KMn_6Bi_5$, pointing to a second-order-effect consequence of SDW. One possible scenario may relate CDW to magnetostriction effects between neighboring magnetic moments which results in a charge modulation period equals half of the spin modulation period in the real space since it does not depend on the directions of those neighboring moments [50]. An incommensurate superspace-group structure refinement is essential to resolve the detailed charge modulation, which, however, cannot be carried out due to the limited quality of the current dataset.

**Conclusions**

Using a combination of neutron and synchrotron x-ray scattering, we successfully determined the coexistence of complex incommensurate charge and spin density waves in KMn$_6$Bi$_5$. Out-of-phase incommensurate SDW with sizeable magnitudes of about 2.46 $\mu_B$ propagate through the Mn pentagons along the *b*-axis direction. In contrast, waves with a small magnitude of no more than 0.29 $\mu_B$ pass through the central Mn chain. The magnetism in KMn$_6$Bi$_5$ has the dichotomy of both local and itinerant characteristics. AFM interactions dominate both within the Mn pentagons and out of the pentagons along the waves in KMn$_6$Bi$_5$. The Mn-cluster chains are antiferromagnetically and ferromagnetically coupled along the *a* and *c* axis, respectively. CDW is much more subtle and probably originates from the magnetostriction effect by SDW due to strong coupling between them [50]. KMn$_6$Bi$_5$, as a unique Q1D system, demonstrates strong interactions among different degrees of freedom such as spin, charge, and lattice. The intertwined orders, including SDW, its induced CDW at ambient pressure and superconductivity under high pressure in KMn$_6$Bi$_5$ provide a fascinating playground to study multiple novel quantum phases and their intimate interplay with the potential for complicated landscapes as in the superconducting cuprates [51].

**Acknowledgements**

This work was supported by the U.S. Department of Energy, Office of Science, Basic Energy Sciences, Materials Sciences and Engineering Division. This research used resources of the Advanced Photon Source; a U.S. Department of Energy (DOE) Office of Science User Facility operated for the DOE Office of Science by Argonne National Laboratory under Contract No. DE-AC02-06CH11357. This research used resources at the High Flux Isotope Reactor, the DOE Office of Science User Facility operated by ORNL. The authors thank Prof. Guang-Han Cao for insightful discussions and Dr. Daniel Phelan for reading and commenting on the manuscript.

**References**

[1]  J. Bardeen, L. N. Cooper, and J. R. Schrieffer, *Theory of Superconductivity*, Physical Review **108**, 1175 (1957).

[2]  M. R. Norman, *The Challenge of Unconventional Superconductivity*, Science (1979) **332**, 196 (2011).

[3]  B. Keimer, S. A. Kivelson, M. R. Norman, S. Uchida, and J. Zaanen, *From Quantum Matter to High-Temperature Superconductivity in Copper Oxides*, Nature **518**, 179 (2015).

[4]  J. Paglione and R. L. Greene, *High-Temperature Superconductivity in Iron-Based Materials*, Nature Physics **6**, 645 (2010).

[5]  N. D. Mathur, F. M. Grosche, S. R. Julian, I. R. Walker, D. M. Freye, R. K. W. Haselwimmer, and G. G. Lonzarich, *Magnetically Mediated Superconductivity in Heavy Fermion Compounds*, Nature **394**, 39 (1998).

[6]  D. Jaccard, H. Wilhelm, D. Jérome, J. Moser, C. Carcel, and J. M. Fabre, *From Spin-Peierls to Superconductivity: $(TMTTF)_2PF_6$ under High Pressure*, Journal of Physics: Condensed Matter **13**, L89 (2001).

[7]  J.-K. Bao et al., *Unique $[Mn_6Bi_5]^-$ Nanowires in $KMn_6Bi_5$: A Quasi-One-Dimensional Antiferromagnetic Metal*, J Am Chem Soc **140**, 4391 (2018).

[8]  L. Chen, L. Zhao, X. Qiu, Q. Zhang, K. Liu, Q. Lin, and G. Wang, *Quasi-One-Dimensional Structure and Possible Helical Antiferromagnetism of $RbMn_6Bi_5$*, Inorganic Chemistry **60**, 12941 (2021).

[9]  S. Long et al., *Flipping of Antiferromagnetic to Superconducting States in Pressurized Quasi-One-Dimensional Manganese-Based Compounds*, ArXivID:2207.14697 (2022).

[10] Z. Y. Liu et al., *Pressure-Induced Superconductivity up to 9 K in the Quasi-One-Dimensional $KMn_6Bi_5$*, Physical Review Letters **128**, 187001 (2022).

[11] P.-T. Yang, Q.-X. Dong, P.-F. Shan, Z.-Y. Liu, J.-P. Sun, Z.-L. Dun, Y. Uwatoko, G.-F. Chen, B.-S. Wang, and J.-G. Cheng, *Emergence of Superconductivity on the Border of Antiferromagnetic Order in $RbMn_6Bi_5$ under High Pressure: A New Family of Mn-Based Superconductors*, Chin. Phys. Lett. **39**, 067401 (2022).

[12] W. Wu, J. Cheng, K. Matsubayashi, P. Kong, F. Lin, C. Jin, N. Wang, Y. Uwatoko, and J. Luo, *Superconductivity in the Vicinity of Antiferromagnetic Order in CrAs*, Nature Communications **5**, 5508 (2014).

[13] J.-G. Cheng, K. Matsubayashi, W. Wu, J. P. Sun, F. K. Lin, J. L. Luo, and Y. Uwatoko, *Pressure Induced Superconductivity on the Border of Magnetic Order in MnP*, Physical Review Letters **114**, 117001 (2015).

[14] T. L. Hung, C. H. Huang, L. Z. Deng, M. N. Ou, Y. Y. Chen, M. K. Wu, S. Y. Huyan, C. W. Chu, P. J. Chen, and T. K. Lee, *Pressure Induced Superconductivity in MnSe*, Nature Communications **12**, 5436 (2021).

[15] T. Giamarchi, *Theoretical Framework for Quasi-One Dimensional Systems*, Chemical Reviews **104**, 5037 (2004).

[16] J. Voit, *One-Dimensional Fermi Liquids*, Reports on Progress in Physics **58**, 977 (1995).

[17] G. Grüner, *The Dynamics of Charge-Density Waves*, Reviews of Modern Physics **60**, 1129 (1988).

[18] J. P. Pouget, S. Ravy, and B. Hennion, *The Charge-Density Wave Instability in Quasi One-Dimensional Conductors*, Phase Transitions **30**, 5 (1991).

[19] D. Jérome, *The Physics of Organic Superconductors*, Science (1979) **252**, 1509 (1991).

[20] G. Grüner, *The Dynamics of Spin-Density Waves*, Reviews of Modern Physics **66**, 1 (1994).

[21] Y. Zhou, C. Cao, and F.-C. Zhang, *Theory for Superconductivity in Alkali Chromium Arsenides A2Cr3As3 (A = K, Rb, Cs)*, Science Bulletin **62**, 208 (2017).

[22] X. Wu, F. Yang, C. Le, H. Fan, and J. Hu, *Triplet $p_z$-Wave Pairing in Quasi-One-Dimensional $A_2Cr_3As_3$ Superconductors (A = K, Rb, Cs)*, Physical Review B **92**, 104511 (2015).

[23] J. Yang, J. Luo, C. Yi, Y. Shi, Y. Zhou, and G. Zheng, *Spin-Triplet Superconductivity in $K_2Cr_3As_3$*, Science Advances **7**, eabl4432 (2021).

[24] K. M. Taddei et al., *Gapless Spin-Excitations in the Superconducting State of a Quasi-One-Dimensional Spin-Triplet Superconductor*, ArXiv:2206.11757 (2022).

[25] S. Reja and S. Nishimoto, *Triplet Superconductivity in Coupled Odd-Gon Rings*, Scientific Reports **9**, 2691 (2019).

[26] J.-K. Bao et al., *Superconductivity in Quasi-One-Dimensional $K_2Cr_2As_3$ with Significant Electron Correlations*, Physical Review X **5**, 011013 (2015).

[27] Y. Zhou, L. Chen, G. Wang, Y.-X. Wang, Z.-C. Wang, C.-C. Chai, Z.-N. Guo, J.-P. Hu, and X.-L. Chen, *A New Superconductor Parent Compound $NaMn_6Bi_5$ with Quasi-One-Dimensional Structure and Lower Antiferromagnetic-Like Transition Temperatures*, Chinese Physics Letters **39**, 047401 (2022).

[28] H. J. Jung, J.-K. Bao, D. Y. Chung, M. G. Kanatzidis, and V. P. Dravid, *Unconventional Defects in a Quasi-One-Dimensional $KMn_6Bi_5$*, Nano Letters **19**, 7476 (2019).

[29] *See Supplemental Material at [URL Will Be Inserted by Publisher] for Additional Information on Experimental Details, Neutron Powder Diffraction, and Synchrotron X-Ray Single Crystal Diffraction*.

[30] H. Cao et al., *DEMAND, a Dimensional Extreme Magnetic Neutron Diffractometer at the High Flux Isotope Reactor*, Crystals (Basel) **9**, 5 (2018).

[31] K. Momma and F. Izumi, VESTA 3 *for Three-Dimensional Visualization of Crystal, Volumetric and Morphology Data*, Journal of Applied Crystallography **44**, 1272 (2011).

[32] J. Angelkort, A. Wölfel, A. Schönleber, S. van Smaalen, and R. K. Kremer, *Observation of Strong Magnetoelastic Coupling in a First-Order Phase Transition of CrOCl*, Physical Review B **80**, 144416 (2009).

[33] T. Kimura, Y. Tomioka, A. Asamitsu, and Y. Tokura, *Anisotropic Magnetoelastic Phenomena in Layered Manganite Crystals: Implication of Change in Orbital State*, Physical Review Letters **81**, 5920 (1998).

[34] N. Narayanan et al., *Magnetic Structure and Spin Correlations in Magnetoelectric Honeycomb $Mn_4Ta_2O_9$*, Physical Review B **98**, 134438 (2018).

[35] T. Chatterji, *Neutron Scattering from Magnetic Materials* (Elsevier, Amsterdam, 2006).

[36] M. Fujihala et al., *Spin Dynamics and Magnetic Ordering in the Quasi-One-Dimensional S=1/2 Antiferromagnet $Na_2CuSO_4Cl_2$*, Physical Review B **101**, 024410 (2020).

[37] S. K. Karna et al., *Helical Magnetic Order and Fermi Surface Nesting in Noncentrosymmetric ScFeGe*, Physical Review B **103**, 014443 (2021).

[38] A. Wills, *Magnetic Structures and Their Determination Using Group Theory*, Le Journal de Physique IV **11**, Pr9 (2001).

[39] E. Fawcett, *Spin-Density-Wave Antiferromagnetism in Chromium*, Reviews of Modern Physics **60**, 209 (1988).

[40] G. Grüner, *Density Waves in Solids*, Vol. CRC Press (CRC Press, 2018).

[41] K. H. Whitmire, *Encyclopedia of Inorganic and Bioinorganic Chemistry* (n.d.).

[42] A. Pandey, P. Miao, M. Klemm, H. He, H. Wang, X. Qian, J. W. Lynn, and M. C. Aronson, *Correlations and Incipient Antiferromagnetic Order within the Linear Mn Chains of Metallic $Ti_4MnBi_2$*, Physical Review B **102**, 014406 (2020).

[43] T. Yamada, N. Kunitomi, Y. Nakai, D. E. Cox, and G. Shirane, *Magnetic Structure of α-Mn*, Journal of the Physical Society of Japan **28**, 615 (1970).

[44] W. Bao, C. Broholm, S. A. Carter, T. F. Rosenbaum, G. Aeppli, S. F. Trevino, P. Metcalf, J. M. Honig, and J. Spalek, *Incommensurate Spin Density Wave in Metallic $V_{2-y}O_3$*, Physical Review Letters **71**, 766 (1993).

[45] Y. Feng et al., *Incommensurate Antiferromagnetism in a Pure Spin System via Cooperative Organization of Local and Itinerant Moments*, Proceedings of the National Academy of Sciences **110**, 3287 (2013).

[46] S. Calder, B. Saparov, H. B. Cao, J. L. Niedziela, M. D. Lumsden, A. S. Sefat, and A. D. Christianson, *Magnetic Structure and Spin Excitations in $BaMn_2Bi_2$*, Physical Review B **89**, 064417 (2014).

[47] Y. Hu et al., *Real-Space Observation of Incommensurate Spin Density Wave and Coexisting Charge Density Wave on Cr (001) Surface*, Nature Communications **13**, 445 (2022).

[48] J. M. Tranquada, *Cuprate Superconductors as Viewed through a Striped Lens*, Advances in Physics **69**, 437 (2020).

[49] J. Zhang et al., *Intertwined Density Waves in a Metallic Nickelate*, Nature Communications **11**, 6003 (2020).

[50] G. van der Laan and I. H. Munro, *X-Ray Scattering and Absorption by Magnetic Materials. By S. W. Lovesey and S. P. Collins (Oxford Series on Synchrotron Radiation, No. 1). Pp. 390. Clarendon Press, 1996. Price (Hardback) £70.00. ISBN 0-19-851737-8.*, Journal of Synchrotron Radiation **5**, 1181 (1998).

[51] E. Fradkin, S. A. Kivelson, and J. M. Tranquada, *Colloquium : Theory of Intertwined Orders in High Temperature Superconductors*, Reviews of Modern Physics **87**, 457 (2015).

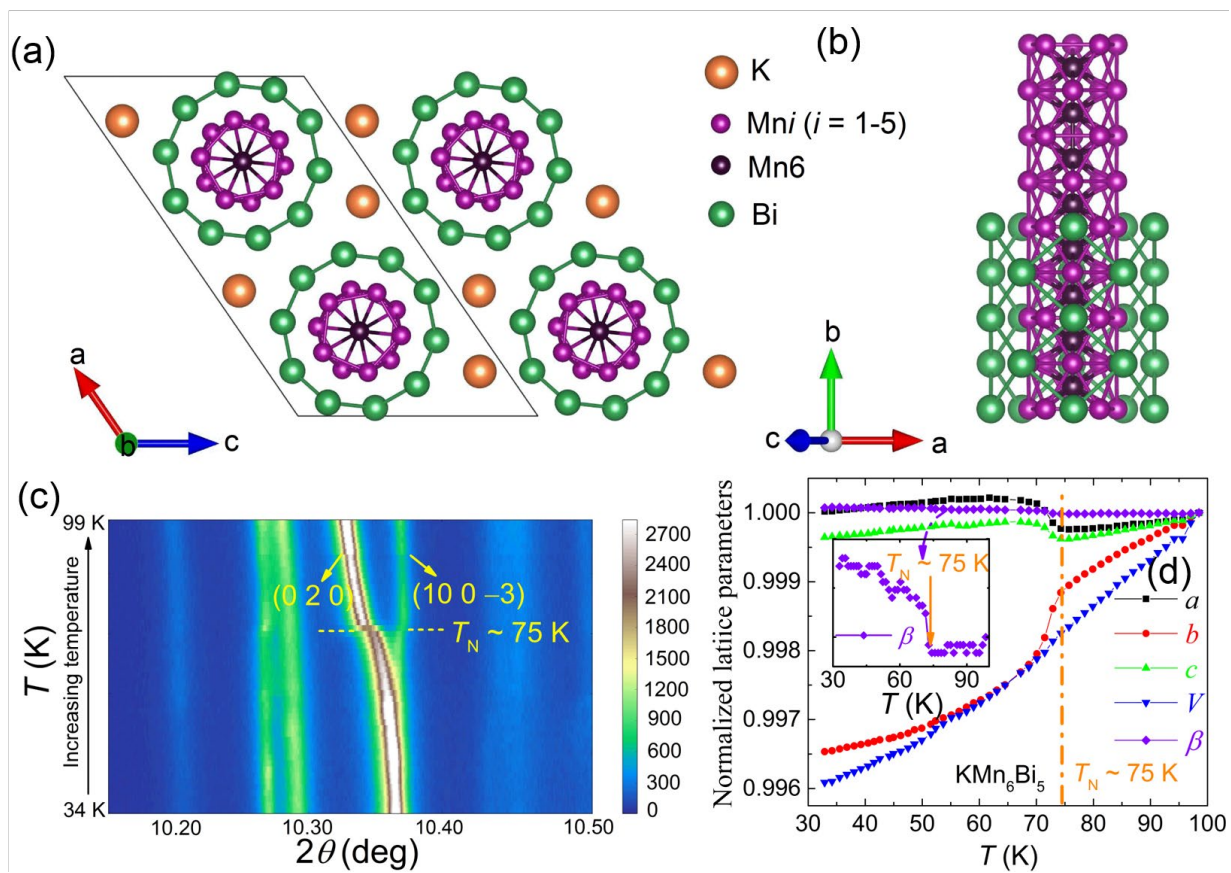

Figure 1 (a) Top view of the crystal structure of KMn$_6$Bi$_5$ down the *b* axis. The bonds were added to help present the nanowire structure of [Mn$_6$Bi$_5$]$^-$ more clearly. (b) The structural configuration of the Bi nanotube and the Mn cluster chain in the [Mn$_6$Bi$_5$]$^-$ nanowire along the *b*-axis direction. (c) Temperature-dependent powder x-ray diffraction of KMn$_6$Bi$_5$ in a selected range of diffraction angles. (d) Temperature-dependent normalized lattice parameters of KMn$_6$Bi$_5$. The inset is a magnified view around the phase transition of the monoclinic angle $\beta$. The orange dash dot line is a guide to the eye.

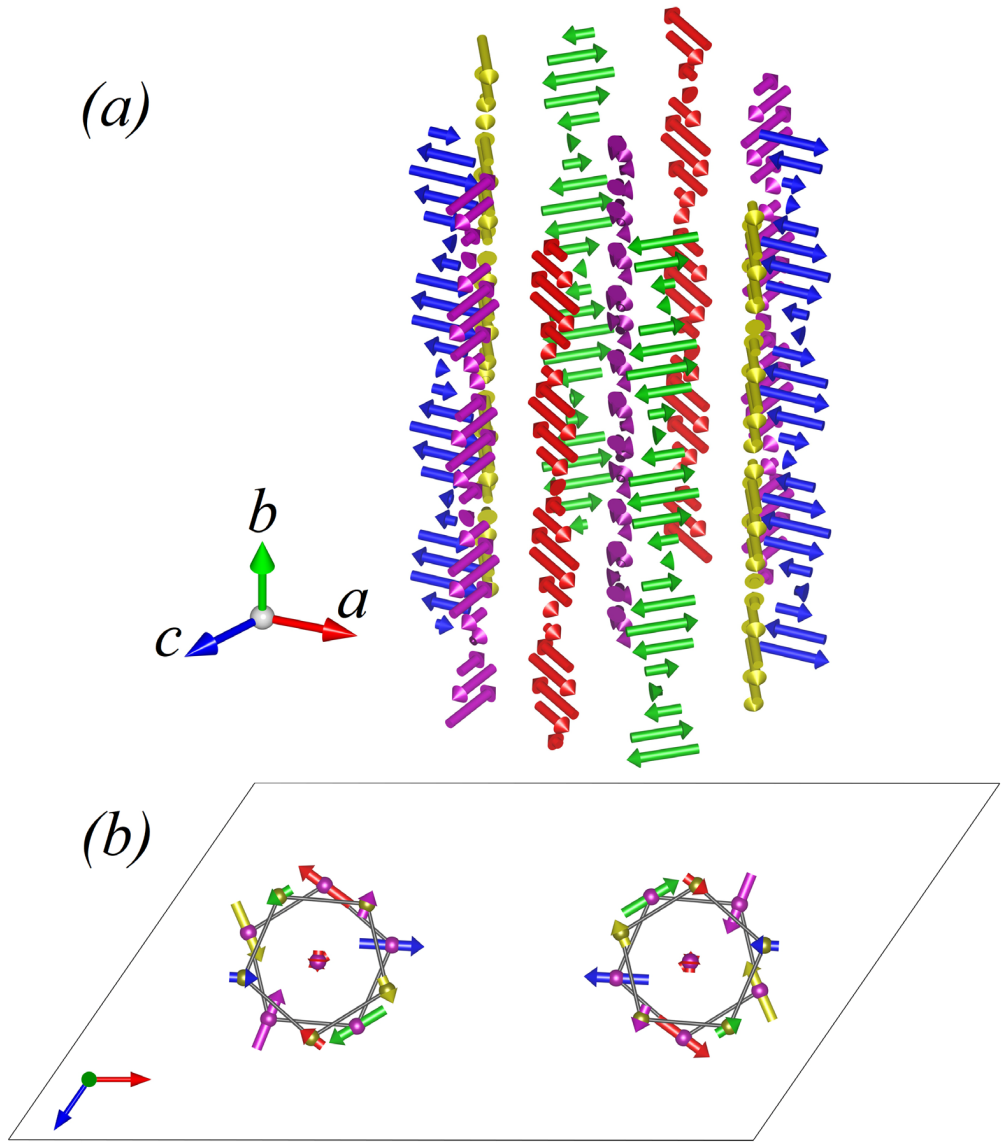

Figure 2 (a) Incommensurate spin density wave (ISDW) magnetic structure of $KMn_6Bi_5$. The structure is plotted over 25-unit cells stacked along the *b*-axis. The unit cell's *b*-parameter is significantly compressed to help visualize the waves over many unit cells. Waves passing through both Mn pentagons are shown together with the chain Mn (purple) through the pentagons' axis. (b) A [010] view of the base unit cell showing the two Mn pentagons (purple spheres at y = 0; dark yellow spheres at y = 0.5). Arrows of the same color are located on two Mn sites that are symmetry related by *C*-centering. Detailed views are shown in Figure 3. See text for more details.

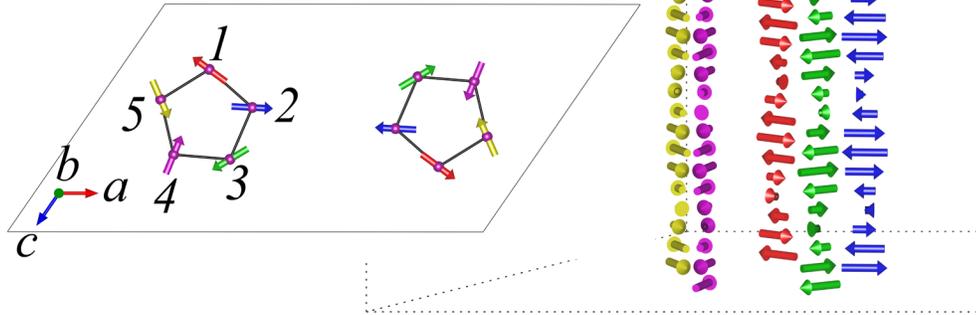

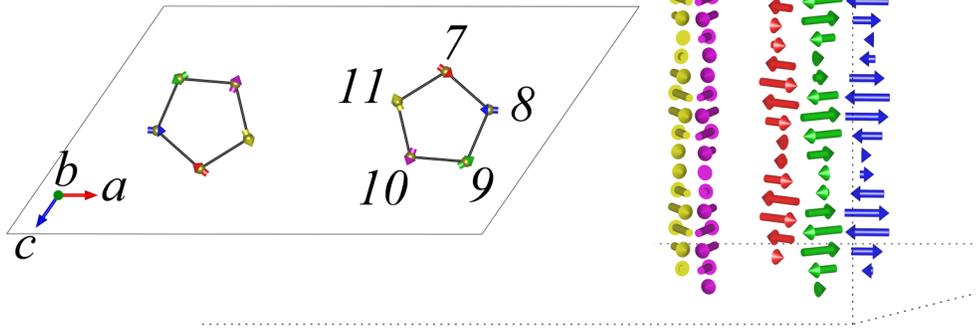

Figure 3 ISDWs passing through (a) the first Mn pentagon located at *y* = 0 as shown to the left of the figure and (b) the second pentagon located at *y* = 0.5. Notice the AFM coupling along the wave propagation direction and that the sublattices are being out-of-phase (See text for more details).

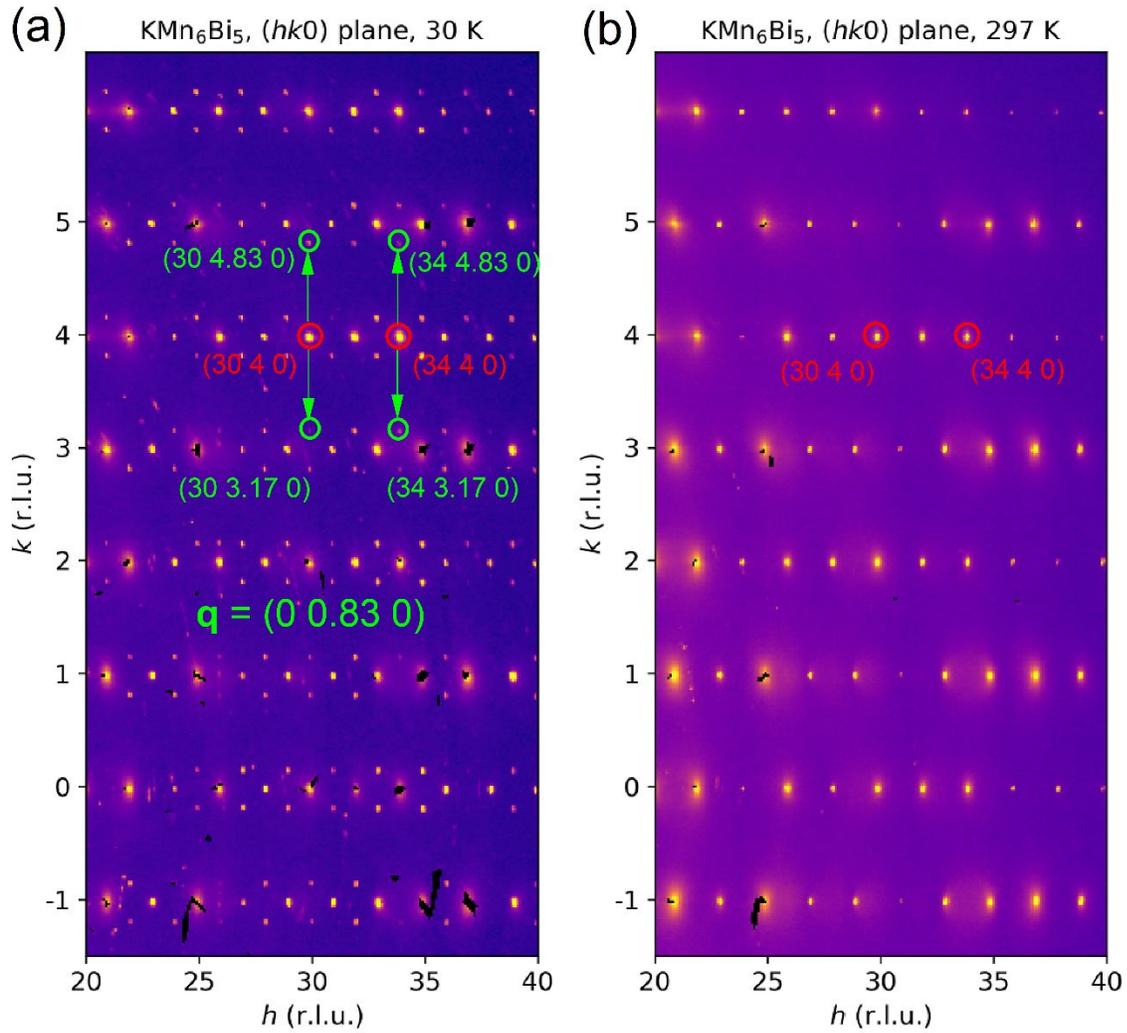

Figure 4. Integrated synchrotron x-ray diffraction intensities measured in the ($hk$0) plane for KMn$_6$Bi$_5$ at (a) 30 K and (b) 297 K, respectively. Examples of CDW satellite peaks and their corresponding **q**-vector are present.

# Supplemental Material

**Experimental Methods**

Single crystals of KMn$_6$Bi$_5$ were grown using excess flux of bismuth as described by Bao et al. [1]. KMn$_6$Bi$_5$ polycrystalline samples were first prepared by a conventional solid-state reaction. A precursor with the composition "Mn$_6$Bi$_5$" was synthesized by heating a mixture of Mn and Bi powders at 673 K for 24 h in an evacuated fused-silica tube. The precursor was then ground into fine powder, mixed with small pieces of K metal in the nominal composition ratio of K$_{1.03}$Mn$_6$Bi$_5$, loaded into an alumina crucible and sintered at 573 K for 12 h in an evacuated fused-silica tube. The sintered mixture was thoroughly ground and reannealed at 803 K for 24 h. The same procedure was repeated at 743 and 658 K to ensure the high-quality of a homogeneous KMn$_6$Bi$_5$ polycrystalline phase.

For synchrotron powder x-ray diffraction, powder sample of KMn$_6$Bi$_5$ was mixed with amorphous SiO$_2$ in the molar ratio of 1:25 and sealed in an evacuated quartz capillary. This diluting step was necessary to minimize the sample's high x-ray absorption. Temperature-dependent data, between 34 and 99 K, were collected while ramping at a constant rate at beamline 11-BM-B of the Advanced Photon Source at Argonne National Laboratory (ANL). Measurements were performed with a radiation wavelength of 0.41 Å. Repeated scans of the same angular range were continuously carried out at an interval of 1.5 K. Contour plots of the temperature-dependent data were created using the software package GSAS-II [2]. Lattice parameters were extracted from Rietveld refinements using the software package suite GSAS/EXPGUI [3,4].

Large three-dimensional volumes of single crystal diffraction were measured at beamline 6-ID-D of the Advanced Photon Source at 30 and 297 K. Data were collected using an x-ray energy of 87 keV with the sample continuously rotating over 360° at a rate of 1 degree per second while a Pilatus 2M CdTe area detector collected images at a 10 Hz rate. The temperature was controlled with an N-Helix Cryocooler at 30 K while the sample was in a dry N-stream at RT.

Magnetic structure determination and order parameter scans were performed using neutron powder diffraction at beamline HB-2A and neutron single crystal diffraction at beamline HB-3A [5] at the high flux isotope reactor (HFIR) at Oak Ridge National Laboratory (ORNL). Integrated intensities of 344 nuclear and 238 magnetic reflections were collected at 4 K using a two-dimensional camera "Anger" [5] that allows for the rapid registration and integration of multiple rocking curves. Solution of the incommensurate magnetic structure was obtained using BasIreps [6] together with the simulated annealing procedure and Rietveld refinement methods as implemented in Fullprof [7].

**Magnetic Structure Modeling**

Analysis of the magnetic reflections collected at 4 K reveals the breakdown of the *C*-centering of the host nuclear lattice for a magnetic structure although no strong reflections violating the centering have been observed because symmetry constraints imposed on the Mn pentagons by a *C*-centered magnetic lattice (six independent sites) result in unsatisfactory agreement factors on the order of ~27%. Removing the *C*-centered magnetic cell results in twelve independent Mn sites and the refinements quickly converged to a solution of an antiferromagnetic spin density wave with the propagation vector along the *b*-axis with a goodness-of-fit agreement factor of about 8%, see Figure S2(b).

The theoretical models [8] of possible commensurate magnetic structures were constructed with significant constraints imposed suggesting that the magnetic structure is likely helical with a propagation vector in the direction of the *b*-axis. In contrast, our model was determined with the Mn magnetic moments free to orient in any direction within the *ac* planes. Canting of the moments in the direction of the *b*-axis was attempted but ultimately dismissed because the angles that resulted from the refinements were small within two or three standard deviations for the current dataset to resolve. Refinements using spiral helical magnetic structures were not successful. Elliptical helical models, on the other hand, always converged to our final solution of a spin-density-wave structure.

When allowed to refine independently, maximum wave amplitudes of ~2.45 to 2.55 $\mu_B$ were obtained for the Mn atoms making the pentagons. The central Mn atoms (*i.e.*, Mn6 and Mn12 sandwiched between two different pentagons) have small, refined amplitudes of less than 0.5 $\mu_B$ with relatively large standard deviations because their contributions to the magnetic scattering are much smaller than the Mn atoms in the pentagons. The magnetic moments of the central Mn atoms are favored to be aligned in the *ac* plane as the case in the Mn pentagons. The magnetic moment size for each Mn atoms in the pentagon remains more or less the same. In the final refinements, the amplitudes of the five pentagon Mn atoms were constrained to be all equal and their phases are set to the same within one pentagon. Despite the constrained amplitudes, the magnetic moments were free to rotate in any direction within the *ac* planes. The breaking of C-centering does not restrict the directions of magnetic moments of two independent pentagons to be symmetry-related. The refinement always leads to almost a colinear configuration for the symmetry-related Mn atoms from C-centering. In our final model, we restricted those magnetic moments and the agreement factor does not increase. The net magnetic moments (~0.3 $\mu_B$/pentagon as the amplitude) from the Mn pentagons in SDWs are not totally cancelled out.

**Table S1** Refined parameters of the nuclear structure at 4 K for a KMn$_6$Bi$_5$ single crystal. Monoclinic space group C2/m with lattice parameters $a$ = 22.8560 Å, $b$ = 4.5565 Å, $c$ = 13.3193 Å, $\alpha$ = 90°, $\beta$ = 124.55°, $\gamma$ = 90°. Unit cell volume = 1168.7449 Å$^3$. Number of effective reflections with $I$ > 1.00 sigma: 344. Residual agreement factors: $R_I$ = 2.25%, $wR_I$ = 3.45%, $R_F$ = 1.90% and $\chi^2$: 3.87.

| Atom | Wyckoff Position | x | y | z | B (Å$^2$) |
| --- | --- | --- | --- | --- | --- |
| K | 4i | 0.8689(2) | 0 | 0.1148(3) | 0.68(8) |
| Mn1 | 4i | 0.1884(2) | 0 | 0.2868(3) | 0.40(8) |
| Mn2 | 4i | 0.3341(2) | 0 | 0.4540(3) | 0.42(8) |
| Mn3 | 4i | 0.6375(2) | 0 | 0.3184(3) | 0.24(7) |
| Mn4 | 4i | 0.2373(2) | 0 | 0.6611(3) | 0.25(7) |
| Mn5 | 4i | 0.1299(2) | 0 | 0.4189(3) | 0.35(8) |
| Mn6 | 4f | 0.25 | 0.25 | 0.5 | 0.30(8) |
| Bi1 | 4i | 0.63406(8) | 0 | 0.1069(1) | 0.25(4) |
| Bi2 | 4i | 0.09466(8) | 0 | 0.5866(1) | 0.29(4) |
| Bi3 | 4i | 0.04062(8) | 0 | 0.1637(1) | 0.29(4) |
| Bi4 | 4i | 0.27714(8) | 0 | 0.2063(1) | 0.32(4) |
| Bi5 | 4i | 0.52477(7) | 0 | 0.3447(1) | 0.23(4) |

**Table S2** SDW magnetic model for KMn$_6$Bi$_5$ at 4 K. The propagation vector (-1, -0.418, 0) is used for the refinements. The magnetic moments on the Mn sites are described by a spherical coordination (R, θ, φ). The *x*-axis is along the *a*-axis and the *y* axis is along the *b*-axis of the crystal structure. The *z*-axis is determined by *x* and *y* axes using the right-hand rule. The magnetic moments are in the *ac* plane (φ = 0). The crystal structure symmetry and its corresponding magnetic symmetry are as follows:
SYMM X, Y, Z MSYM u,v,w, 0.00
SYMM -X+1,Y,-Z+1 MSYM -u,v,-w, 0.00

| Atom | Amplitude (R) (μ$_B$) | Theta (θ) (degree) | Phi (φ) (degree) | Phase of waves (2π) |
|---|---|---|---|---|
| Mn1 | 2.46 | -126 | 0 | 0 (fixed) |
| Mn2 | 2.46 | 87 | 0 | 0 (fixed) |
| Mn3 | 2.46 | -59 | 0 | 0 (fixed) |
| Mn4 | 2.46 | 158 | 0 | 0 (fixed) |
| Mn5 | 2.46 | 24 | 0 | 0 (fixed) |
| Mn6 | 0.29 | -58 | 0 | 0.96 |
| Mn7 | 2.46 | -126 | 0 | 0.29 |
| Mn8 | 2.46 | 87 | 0 | 0.29 |
| Mn9 | 2.46 | -59 | 0 | 0.29 |
| Mn10 | 2.46 | 158 | 0 | 0.29 |
| Mn11 | 2.46 | 24 | 0 | 0.29 |
| Mn12 | 0.29 | -183 | 0 | 0.1 |

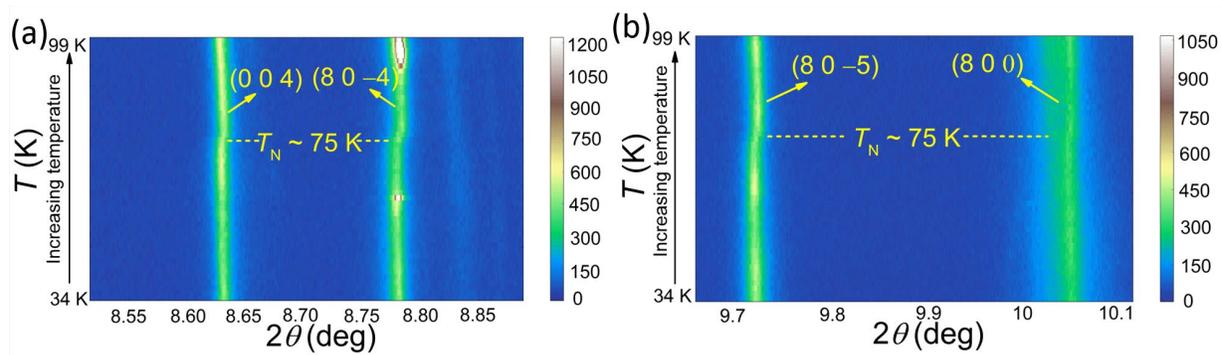

Figure S1 (a) (b) Temperature-dependent powder x-ray diffraction of KMn$_6$Bi$_5$ in a selected range of diffraction angles. The relevant reflections (8 0 -5), (8 0 0), (0 0 4) and (8 0 -4) were tracked across the phase transition $T_N$. The dashed lines are a guide to the eye.

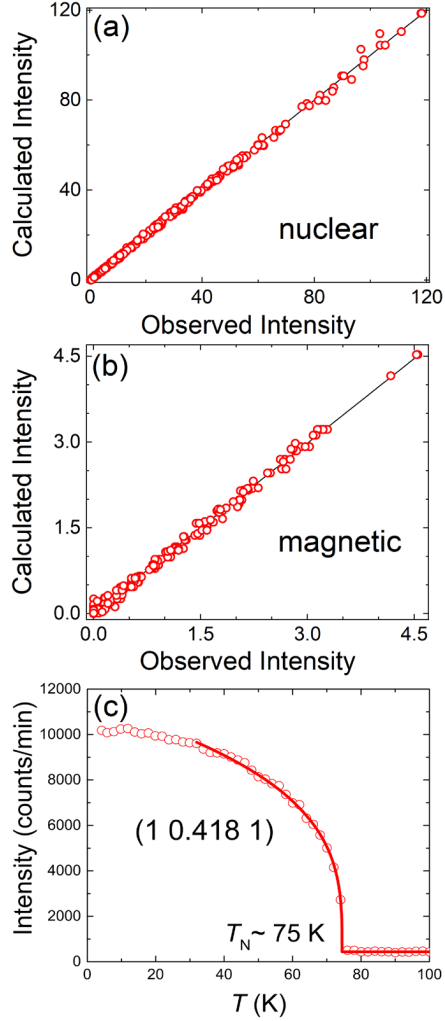

Figure S2 The relationship between the observed and calculated reflection intensity for refined nuclear (a) and magnetic (b) structures from single crystal neutron diffraction. The data points are almost located on the straight line, indicating an excellent fit. (c) Temperature-dependent integrated intensity of the magnetic reflection (1 0.418 1) from single crystal neutron diffraction. The red line is the fit of the critical behavior near the phase transition. It can be well fitted using a standard power-law equation $I = I_0 + I_M(1 - \frac{T}{T_N})^{2\beta}$ as described by Chatterji et. al. [9] and Karna et al. [10]. Here, $I_0$ is a temperature-independent term arising mainly from noise background of detector, $I_M$ reflects magnetic contributions at 0 K, $T_N$ is the magnetic phase transition temperature and $\beta$ is the critical exponent. The fitted value of $T_N$ = 74.5(3) K is fully consistent with the anomaly observed in the x-ray diffraction data and the phase transition reported by Bao *et al.* [1].

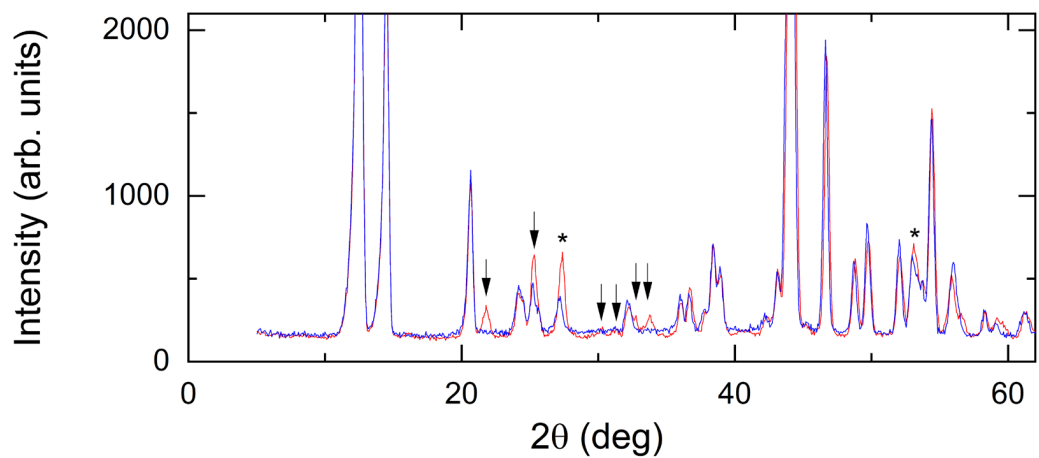

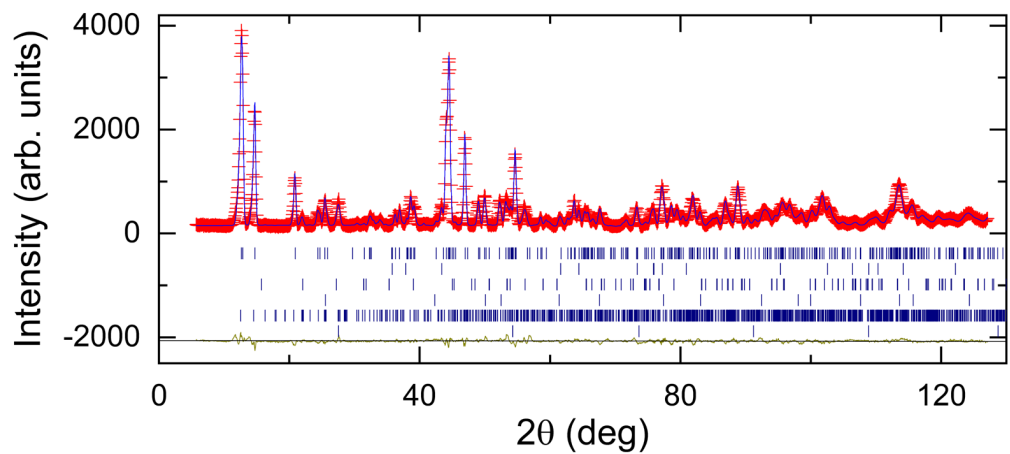

Figure S3. Neutron Powder Diffraction. (a) Raw data collected at 3 K and 297 K, highlighting the magnetic peaks of KMn$_6$Bi$_5$ (arrows) that appear below 75 K. Two magnetic peaks that belong to a tiny impurity phase MnO are also observed (asterisks). (b) Best-fit Rietveld refinements of the nuclear and magnetic structure of KMn$_6$Bi$_5$. Observed intensities, calculated intensities, and the difference plot are represented by plus symbols (+), a solid blue line, and a solid dark yellow line, respectively. Tick marks (from top to bottom) indicate the Bragg positions of the nuclear reflections of KMn$_6$Bi$_5$, Bi, MnO, KBi$_2$, and the magnetic reflections of KMn$_6$Bi$_5$ and MnO. Weight Fractions for KMn$_6$Bi$_5$: 92.8(9)%, Bi: 3.24(2)%, MnO: 1.07(4), KBi$_2$: 2.95(12).

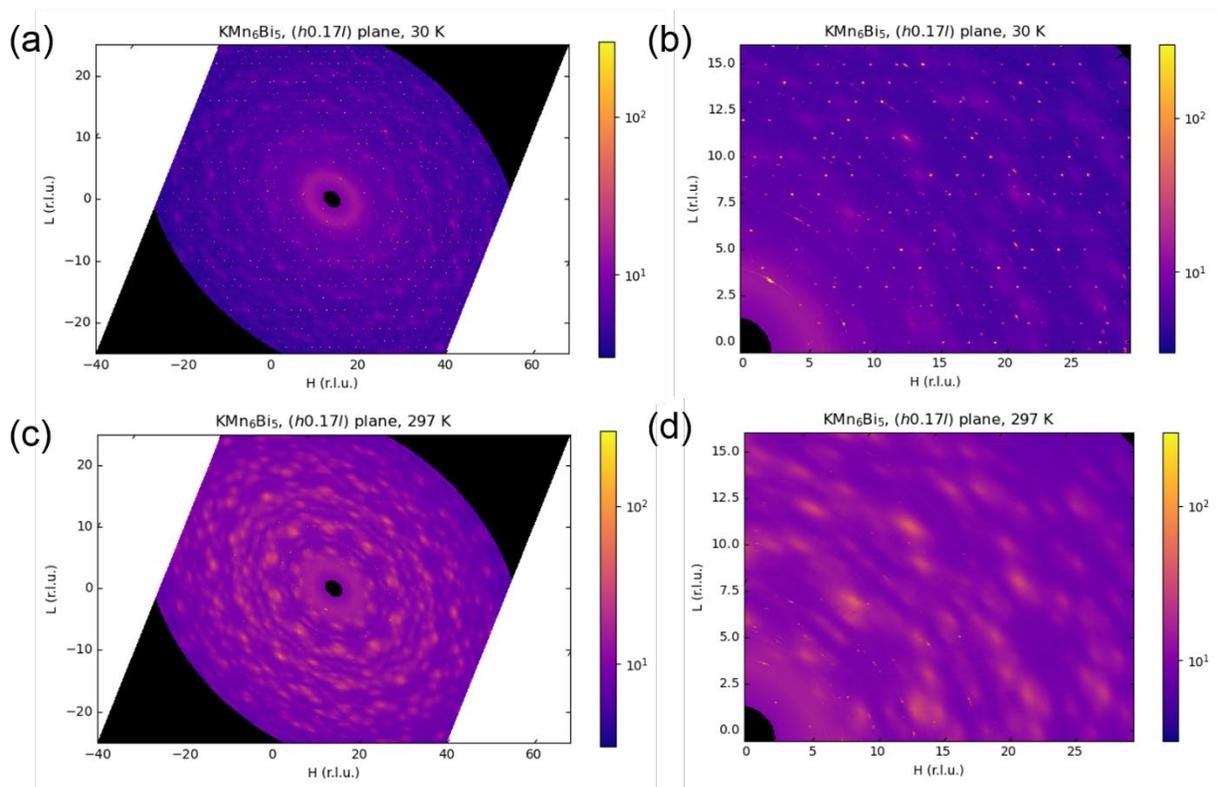

Figure S4 (a), (c) Integrated synchrotron x-ray scattering intensities of ($h$1.17$l$) planes at 30 and 297 K, respectively. (b) and (d) are the corresponding zoomed plots for (a) and (c), respectively. Satellite peaks are well resolved at 30 K in a wide reciprocal space while they are absent at 297 K

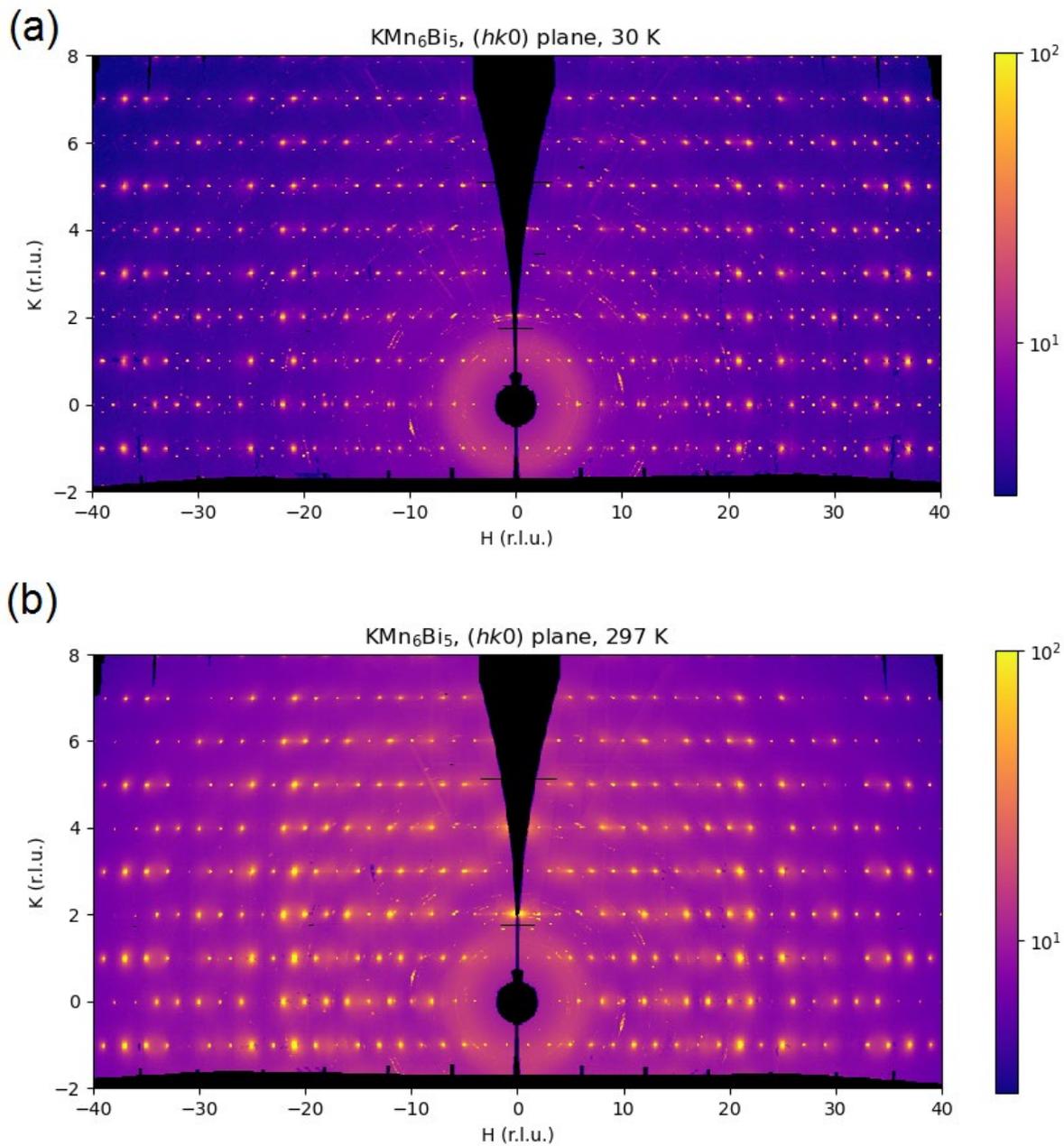

Figure S5 (a), (b) Integrated synchrotron x-ray scattering intensities of (*hk*0) planes in a wide reciprocal space at 30 and 297 K, respectively. The satellite peaks are well resolved at 30 K while they are absent at 297 K.

**Supplemental References**

[1] J.-K. Bao et al., *Unique [Mn$_6$Bi$_5$]$^-$ Nanowires in KMn$_6$Bi$_5$: A Quasi-One-Dimensional Antiferromagnetic Metal*, J Am Chem Soc **140**, 4391 (2018).

[2] B. H. Toby and R. B. von Dreele, *GSAS-II: The Genesis of a Modern Open-Source All Purpose Crystallography Software Package*, Journal of Applied Crystallography **46**, 544 (2013).

[3] Larson A. C. and R. B. von Dreele, *General Structure Analysis System (GSAS); Los Alamos National Laboratory Report LAUR 86-748*, Los Alamos National Laboratory: Los Alamos, NM, (2004).

[4] B. H. Toby, *EXPGUI, a Graphical User Interface for GSAS*, Journal of Applied Crystallography **34**, 210 (2001).

[5] H. Cao et al., *DEMAND, a Dimensional Extreme Magnetic Neutron Diffractometer at the High Flux Isotope Reactor*, Crystals (Basel) **9**, 5 (2018).

[6] J. Rodriguez-Carvajal, *BASIREPS: A Program for Calculating Irreducible Representations of Space Groups and Basis Functions for Axial and Polar Vector Properties*, Solid State Phenom **170**, (2011).

[7] J. Rodríguez-Carvajal, *Recent Advances in Magnetic Structure Determination by Neutron Powder Diffraction*, Physica B: Condensed Matter **192**, 55 (1993).

[8] L. Chen, L. Zhao, X. Qiu, Q. Zhang, K. Liu, Q. Lin, and G. Wang, *Quasi-One-Dimensional Structure and Possible Helical Antiferromagnetism of RbMn$_6$Bi$_5$*, Inorganic Chemistry **60**, 12941 (2021).

[9] T. Chatterji, *Neutron Scattering from Magnetic Materials* (Elsevier, Amsterdam, 2006).

[10] S. K. Karna et al., *Helical Magnetic Order and Fermi Surface Nesting in Noncentrosymmetric ScFeGe*, Physical Review B **103**, 014443 (2021).